\documentclass[12pt,fleqn]{article}

\usepackage{amsmath,amsfonts,bm,amssymb}
\usepackage{amsthm}

\usepackage{eucal}
\usepackage{eufrak}

\usepackage[compat2,a4paper]{geometry}

\usepackage[authoryear]{natbib}
\usepackage{url}

\usepackage{multirow}
\usepackage{rotating}

\usepackage{graphicx}
\graphicspath{{figures/}} 

\DeclareMathOperator{\E}{E}

\renewcommand{\le}{\leqslant}
\renewcommand{\leq}{\leqslant}

\renewcommand{\geq}{\geqslant}
\newcommand{\w}[2]{\ensuremath{w_{#1}^{(#2)}}}
\newcommand{\D}{\displaystyle}
\renewcommand{\r}[2]{\ensuremath{r_{#1}^{(#2)}}}
\newcommand{\eqGaussian}{\stackrel{\text{G}}{=}}

\newcommand{\remark}[1]{\marginpar{\scriptsize#1}}
\renewcommand{\remark}[1]{\marginpar{}}   

\newcommand{\captionfonts}{\footnotesize}

\makeatletter  
\long\def\@makecaption#1#2{%
  \vskip\abovecaptionskip
  \sbox\@tempboxa{{\captionfonts #1: #2}}%
  \ifdim \wd\@tempboxa >\hsize
    {\captionfonts #1: #2\par}
  \else
    \hbox to\hsize{\hfil\box\@tempboxa\hfil}%
  \fi
  \vskip\belowcaptionskip}
\makeatother   

\begin{document}
 

\title{\LARGE The Reality Game\footnote{
We would like to thank Jonathan Goler for his help with simulations
in the early days of this project,  Michael Miller for
reproducing the results and clarifying the scaling in
Figure~\ref{efficiency}, and Brian Arthur, Larry Blume, Giovanni Dosi and Ole Peters for useful comments.  JDF would like to thank Barclays Bank, Bill Miller, and National Science Foundation grant 0624351 for support.  Any opinions, findings, and conclusions or recommendations expressed in this material are those of the authors and do not necessarily reflect the views of the National Science Foundation. }
}

\author{Dmitriy Cherkashin\,$^{\textrm{a},\textrm{b}}$
 \and
J. Doyne Farmer\,$^{\textrm{b},\textrm{c}}$\footnote{Corresponding author, jdf@santafe.edu} \and
Seth Lloyd\,$^{\textrm{d},\textrm{b}}$}

\date{}
\maketitle
\small
\begin{center}
  $^\textrm{a}$~\emph{Mathematics Dept., University of Chicago, Chicago, IL 60637}\\
  $^\textrm{b}$~\emph{Santa Fe Institute, 1399 Hyde Park Rd., Santa Fe NM 87501}\\
  $^\textrm{c}$~\emph{LUISS Guido Carli, Viale Pola 12, 00198, Roma, Italy}\\
  $^\textrm{d}$~\emph{Mechanical Engineering Dept., MIT, 77 Massachusetts Ave., Cambridge MA 02139-4307}
\end{center}
\normalsize

\vspace{0.3cm}

\begin{abstract}
We introduce an evolutionary game with feedback between perception and reality, which we call the reality game.  It is a game of chance in which the probabilities
for different objective outcomes (e.g., heads or tails in a coin toss) depend on the amount
wagered on those outcomes.  
By varying the `reality map', which relates the amount wagered to the
probability of the outcome, it is possible to move continuously from
a purely objective game in which probabilities have no dependence on
wagers to a purely subjective game in which probabilities equal the
amount wagered.  We study self-reinforcing games, in which betting more on an outcome increases its odds, and self-defeating games, in which the opposite is true. 
This is investigated in and out of equilibrium, with and without rational
players, and both numerically and analytically. We introduce a method of measuring the inefficiency of the
game, similar to measuring the magnitude of the arbitrage opportunities in a financial market.  We prove that the inefficiency converges to equilibrium as a power law with an extremely slow rate of convergence: The more subjective the game, the slower the convergence.
 \end{abstract}

\smallskip

\smallskip


\newpage

$\ldots$ {\it if a dream can tell the future it can also thwart that future.  For God will not permit that we shall know what is to come.  He is bound to no one that the world unfold just so upon its course and those who by some sorcery or by some dream might come to pierce the veil that lies so darkly over all that is before them may serve by just that vision to cause that God should wrench the world from its heading and set it upon another course altogether and then where stands the sorcerer?  Where the dreamer and his dream?

Cormac McCarthy, The Crossing}

\section{Introduction}

\subsection{Motivation}

To motivate the idea that outcomes might depend on subjective
perception \cite{Keynes36} used the metaphor of a beauty contest in
which the goal of the judges is not to decide who is most beautiful,
but rather to guess which contestant will receive the most votes from the other
judges.  Economic problems
typically have both purely objective components, e.g.\ how much
revenue a company creates, as well as subjective components, e.g.\
how much revenue investors collectively \textit{think} it will create.  The
two are inextricably linked.  For example, if investors think a company will succeed they will invest more money, which may allow the company to increase its revenue.  In general subjective perceptions affect investment decisions, which affect objective outcomes, which in turn affect subjective perceptions.  Keynes's model only deals with the purely subjective elements of this problem, in the sense that the choices of the judges do not  change how beautiful the contestants are.  

We are interested in developing a simple conceptual model for the more general case where subjective perception affects objective outcomes.  In our game the opinions of the judges \textit{do} affect the beauty of the contestants.  The importance of this in financial markets has been stressed by \cite{Soros87}, who calls this `market reflexivity'.  Our model takes the form of a simple game of chance in which the probability of outcomes can depend on the amount bet on those outcomes.  This dependence is controlled by the choice of a `reality map' that mediates
between subjective perception and objective outcome.  The form of
the reality map can be tuned to move continuously from purely
objective to purely subjective games, and to incorporate both positive and negative feedback.\footnote{
Some of the results in this paper are
presented in more detail in \cite{Cherkashin04}.}
%

Consider a probabilistic event, such as a coin toss or the outcome
of a horse race.  Now suppose that the odds of the outcomes of the
event depend on the amount wagered on them.  In the case of a coin
toss, this means that the probability of heads is a function (the
reality map) of the amount bet on heads.  For a purely objective
event, such as the toss of a fair coin, the reality map is simple:
the probability of heads is 1/2, independent of the amount bet on it, and
thus the reality map is constant.  But many situations depend on subjective elements.  In the case of a horse race, for instance, a jockey riding a strongly favored horse may
make more money if he secretly bets on the second most favored horse
and then intentionally loses the race.  This is an example of a
self-defeating reality map: if jockeys misbehave, then as the horse becomes
more popular, the objective probability that it will win decreases.
Alternatively, in the economic setting we discussed above, if people like growth
strategies then they invest in companies whose prices are going up,
which in turn drives prices further up.  This is an example of a
self-reinforcing reality map.  Our model makes it possible to qualitatively study these diverse cases within the context of game theory in a simple and consistent setting.

In this paper we study a game of chance under a variety of different reality maps, ranging from purely objective to purely subjective, and including both self-defeating and self-reinforcing cases.  We study this as an evolutionary game in which players have fixed strategies.  Through time there is a reallocation of wealth toward a single strategy.  By introducing a rational player we can define the inefficiency of the game in a way that is similar to how one might define the inefficiency in a financial market based on the magnitude of the arbitrage opportunities.  We show that the inefficiency converges to zero very slowly, and in particular, when the reality map is close to being subjective, it converges extremely slowly.

\subsection{Review of related work}

We use the framework developed by \cite{Kelly56} and  \cite{CoverThomas:book}.  They study a repeated game of chance in which individual investors use fixed strategies and payoffs are determined according to pari-mutuel betting.   The players do not consume anything and reinvest all their wealth at every step.  They show that the relative performance of strategies can be understood in terms of an entropy measure, and that the strategy that asymptotically accumulates all the wealth is that which maximizes log-returns, as originally shown by \cite{Kelly56} and developed by \cite{Breiman61} and others.  Similar results in a slightly different framework were independently developed by \cite{Blume93}.  The motivation for the Blume and Easley model was to understand whether rationality is the only criterion for survival under natural selection, a topic that has generated considerable interest in the evolutionary economics literature.\footnote{
Some relevant examples in the evolutionary economics literature include \cite{Blume-Easley92,Blume-Easley02}, \cite{Sandroni00}, \cite{Alos-Ferrer05}, \cite{Hens05b}, and \cite{Anufriev06}.  See also the introduction to a special issue on evolutionary economics by \cite{Hens05}.  Note that this line of work has failed to notice earlier work mentioned above from the literature on information theory and the theory of gambling, which we believe would help clarify the debate.}

All of the work above assumes that objective outcomes are exogenously given.  Our results here differ because we extend the framework of \cite{Kelly56} and \cite{CoverThomas:book} to allow objective outcomes to be determined endogenously, i.e.\ we let the objective payoffs change in a way that depends on the bets of the players.  The idea that agent actions can influence objective outcomes is an old one in economics.  Examples include studies of increasing returns (\cite{Arthur94:book}), the El Farol model (\cite{Arthur94:paper}) and its close relative the minority game (\cite{Challet97}), and cobweb models (\cite{Hommes91,Hommes94}).  In such models the feedback between actions and outcomes generates historical dependencies and lock-ins to one of multiple expectational equilibria.  Positive feedback has also been studied in terms of the generalized Polya urn problem; in the simplest case a ball is drawn from an urn and replaced by two balls of the same color, which asymptotically results in lock-in to one color or the other (\cite{Arthur83,Dosi94}).  In a somewhat different vein, a simple example of a game that changes due to the players' behaviors and states was developed by \cite{akiyama-kaneko00}.  The model we introduce here exhibits many of the behaviors seen in these previous models, but has the advantage of being general yet simple, with tunable levels of positive or negative feedback.

\subsection{Outline of paper}

In Section~2 we define the reality game.  In Section~3 we study the dynamics of the objective bias numerically as a function of time under different specifications of the reality map, and in Section~4 we study the wealth dynamics.  In Section~5 we give an analytic explanation for the numerical results, showing how the wealth dynamics and objective bias approach the stable fixed points of the reality game according to a power law. In Section~6 we introduce a notion of a rational player and show how such players can achieve increasing returns.  We compute the equilibria of the game and discuss the circumstances under which myopic optimization is effective.  In Section~7 we introduce a notion of efficiency similar to arbitrage efficiency in financial markets and introduce a quantitative way to measure the inefficiency of the game when it is out of equilibrium.  We study the inefficiency of time both numerically and analytically, and show that its convergence to equilibrium is extremely slow and depends on the degree of subjectivity.  Finally Section~8 contains a few concluding remarks. 

\section{Game definition}

\subsection{Wealth dynamics}

Let $N$ agents place wagers on $L$ possible outcomes.  In the case
of betting on a coin, for example, there are two outcomes, heads and
tails. Let $s_{il}$ be the fraction of the $i$-th player's wealth
$w_i$ that is wagered on the $l$-th outcome.  The vector
$(s_{i1},\dots,s_{iL})$ is the $i$-th player's \emph{strategy},
and $p_{il} = s_{il} w_i$ is the amount of money bet on the $l$-th
outcome by player $i$.  Let $p_l=\sum_i p_{il}$ be the total wager
on the $l$-th outcome.  If the winning outcome is $l =  \lambda$,
the payoff\footnote{
Since each player bets all her wealth the payoff is the same as that player's wealth at the next time step.}
$\pi_i$ to player $i$ is $\pi_{i\lambda}$ which is proportional to the
amount that player bets and inversely proportional to the total
amount everyone bets, i.e.
\begin{equation}
\pi_{i\lambda} = \frac{p_{i\lambda}}{p_\lambda} =
\frac{s_{i\lambda}w_i}{p_\lambda}\,. 
\label{payoff}
\end{equation}
This corresponds to what is commonly called pari-mutuel betting.  We
assume no ``house take'', i.e.\ a fair game.  Assume \ $\sum_l s_{il} =
1$, i.e.\ that each player bets all her money at every iteration of the game, typically betting non-zero amounts on each possible outcome.
The total wealth is conserved and is normalized to sum to one,
\begin{equation}
\sum_i w_i=\sum_{i,l} p_{il} = 1\,. 
\label{wealthnormal}
\end{equation}

We will call $q_l$ the probability of outcome $l$, where $\sum_l q_l
= 1$.  The expected payoff $\E[ \pi_i]$ is
\begin{equation}
\E [\pi_i] = \sum_l q_l \pi_{il}\,.
\label{expectpayoff}
\end{equation}
 If the vector $q$ is fixed, after playing the game repeatedly for $t$ rounds
the wealth updating rule is
\begin{equation}
w_i^{(t+1)}=\frac{s_{i\lambda}w_{i}^{(t)}}{p_\lambda}.
\label{expectwealth}
\end{equation}
As originally pointed out by \cite{CoverThomas:book} and \cite{Blume93}, this is equivalent to Bayesian inference, where the initial wealth
 $w_i^{(t)}$ is interpreted as the prior probability that $q_\lambda =
s_{i\lambda}$ and the final wealth $w_i^{(t+1)}$ is its posterior
probability.  In Bayesian inference, models whose predictions match
the actual probabilities of outcomes accrue higher \textit{a
posteriori} probability as more and more events occur.  Here
players whose strategies more closely match actual outcome
probabilities accrue wealth on average at the expense of players
whose strategies are a worse match.

\subsection{Strategies}

We first study fixed strategies.  For convenience we restrict the
possible number of outcomes to $L = 2$, so that we can think of this
as a coin toss with possible outcomes heads and tails.  Because the
players are required to bet all their money on every round,  $s_{i1}
+ s_{i2} = 1$, we can simplify the notation and let $s_i = s_{i1}$
be the amount player $i$ bets on heads --- the amount bet on tails is determined
 automatically.  Similarly $q = q_1$ and $p = p_1$. The space of possible strategies corresponds to the
unit interval $[0,1]$. We will typically simulate $N$ fixed
strategies, $s_i = i/(N - 1)$, where $i = 0, 1, \ldots, N-1$.
 Later
on we will also add players that are rational in the sense that they know  the strategies of
 all other players and dynamically adapt their own strategies accordingly to maximize
a utility function.

\subsection{Reality maps}

The game definition up to this point follows \cite{CoverThomas:book}.  We generalize their game by
allowing for the possibility that the objective probability $q$ for
heads is not fixed, but rather depends on the net amount $p$ wagered
on it. The \textit{reality map} $q(p)$, where $0 \leq q(p) \leq
1$, fully describes the relation between bets and outcomes.  We
restrict the problem slightly by requiring that $q(1/2) = 1/2$.  We
do this to give the system the chance for the objective outcome, as
manifested by the bias of the coin, to remain constant at $q = 1/2$.
We begin by studying the case where $q(p)$ is a monotonic function,
which is either nondecreasing or nonincreasing.  Letting $q'(p) =
dq/dp$, we distinguish the following possibilities:
\begin{itemize}
\item \textit{Objective.}  $q(p) = 1/2$,  i.e.\ it is a fair coin independent of the amount wagered.  (Other
values of $q = \textit{const}$ are qualitatively similar to $q=1/2$.)
\item \textit{Self-defeating.}  $q'(p) < 0$, e.g.\ $q(p) = 1 - p$. In this case the coin tends to oppose the collective perception, e.g.\ if people collectively bet on heads, the coin is biased toward tails.
\item \textit{Self-reinforcing.}  $q'(p) > 0$. The coin tends to reflect the collective perception, e.g.\ if people collectively bet on heads, the coin becomes more biased toward heads.  A special case of this is \textit{purely subjective}, i.e.\ $q(p) = p$, in which the bias simply reflects people's bets.
\end{itemize}

\begin{figure}[htb]
\begin{center}
\includegraphics[scale = 0.6]{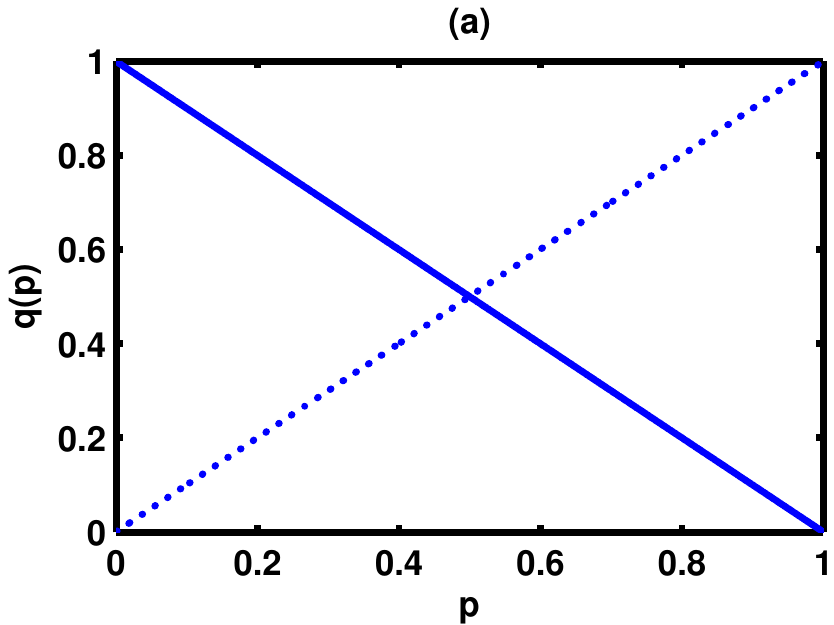}
\includegraphics[scale = 0.6]{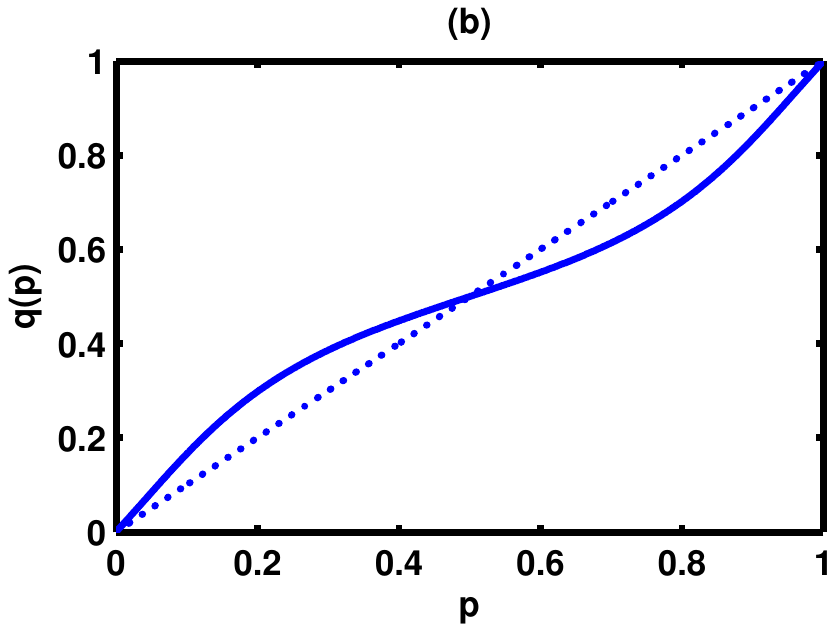}
\includegraphics[scale = 0.6]{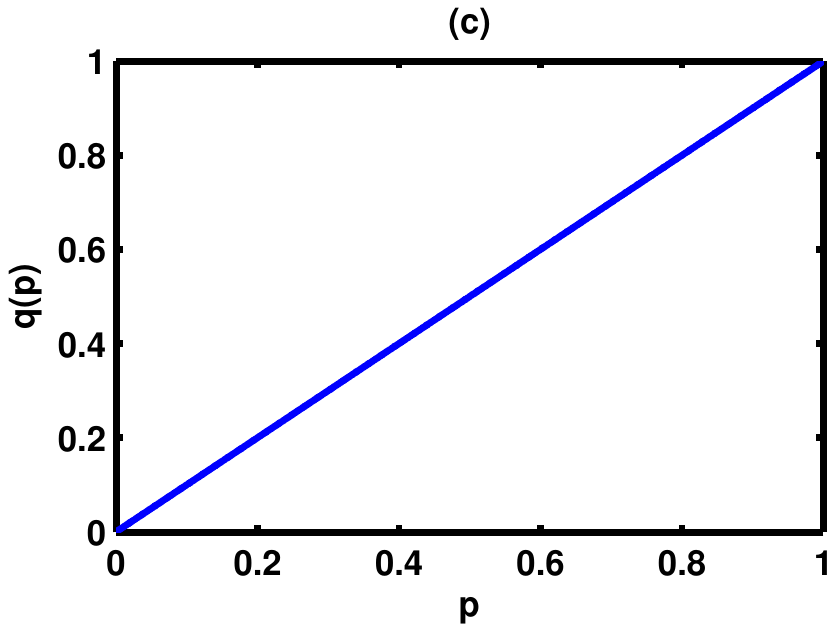}
\includegraphics[scale = 0.6]{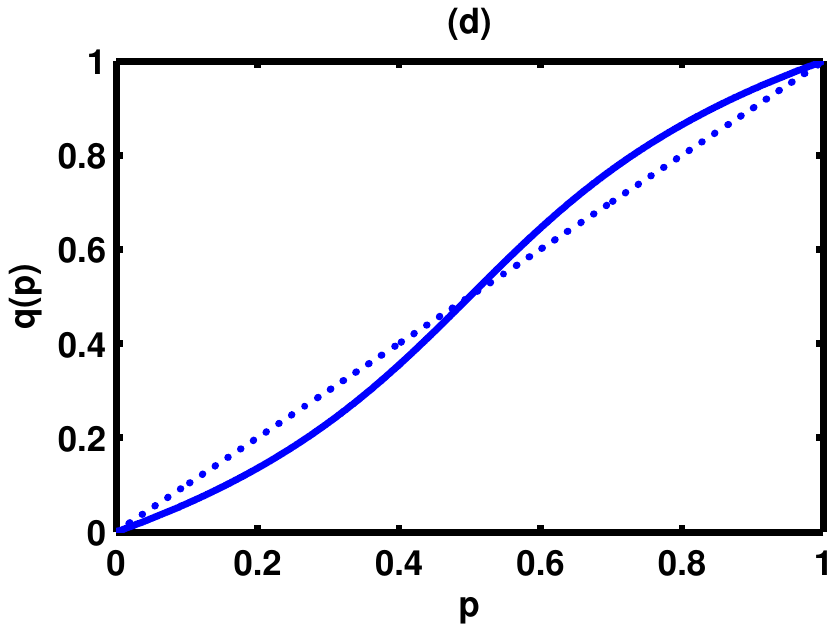}
\includegraphics[scale = 0.6]{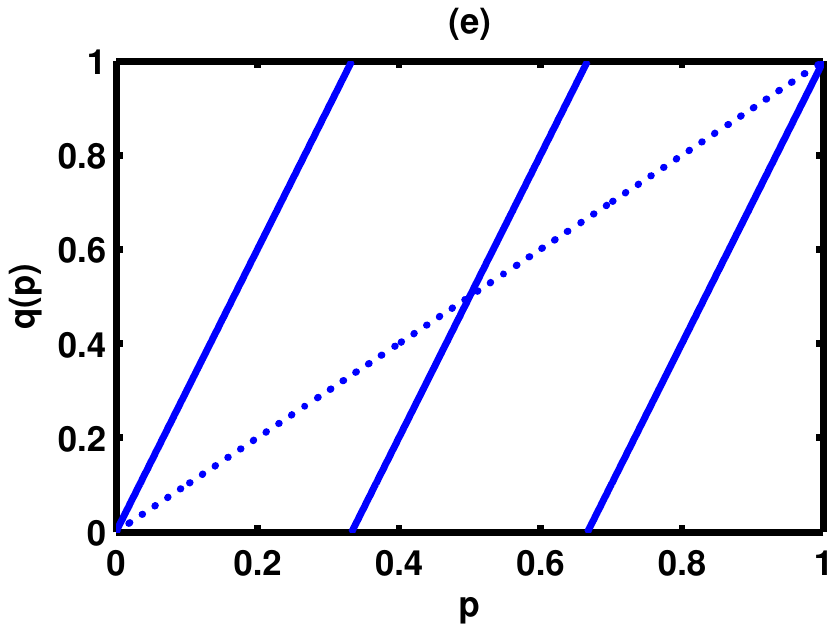}
\caption{In the reality game, the probability $q$ of the outcome
heads is a function of the total fraction of wealth $p$ 
bet on heads.  Figure~(a) shows a purely
self-defeating map in which the probability of the outcome heads 
is negatively correlated with the amount bet on heads.
The only fixed point of this map, corresponding to a possible
equilibrium point for the game, is at $p=1/2$.
Figures (b), (c), (d), show positively reinforcing maps 
where the probability of heads
is positively correlated with the amount bet on heads: here,
$\alpha$ is a parameter that determines the degree of correlation of
the outcome probabilities with the inputs.   In (b) $\alpha = 1/2$; (c) is the identity map; and in (d) $\alpha = 1.5$.  
Finally, figure~(e) shows a multi-modal reality map, $q(p) = 3p \mod 1$,
with three fixed points.}
\label{realityMaps}
\end{center}
\end{figure}

It is convenient to have a one parameter family of reality maps that
allows us to tune from objective to self-reinforcing. We choose the
family
\begin{equation}
\label{qalpha}
q_{\alpha}(p) = \frac{1}{2} + \frac{1}{\pi} \arctan \frac{\pi \alpha
(p-\frac{1}{2})}{1-(2p-1)^2},
\end{equation}
as shown in Figure~\ref{realityMaps}. 
The parameter $\alpha$ is the slope at $p = 1/2$.  When $\alpha =
0$, $q(p)$ is constant (purely objective), and when $\alpha > 0$,
 $q(p)$ is self-reinforcing.  The derivative $q'_\alpha(1/2)$ is an
 increasing function of $\alpha$; when $\alpha = 1$, $q'(1/2) = 1$, and
 $q(p)$ is close to the identity map.\footnote{
An inconvenient aspect of
this family is that $q_{\alpha}(p)$ does not contain the function
$q(p)=p$.  However, $q_1(p)$ is very close to $q(p)=p$ (the
difference does not exceed $0.012$, with the average value less than
 half of this).  Still, to avoid any side effects, we study the purely subjective case using
$q(p) = p$.}
We study the self-defeating case separately using the
map $q(p) = 1 - p$.  Finally, to study a more complicated example, we study the multi-modal reality map, $q(p) = 3p \mod 1$, which has three fixed points.

\section{Dynamics of the objective bias}
\label{sec:dynamics-of-bias}

\begin{figure}[htb]
\begin{center}
\includegraphics[scale = 0.6]{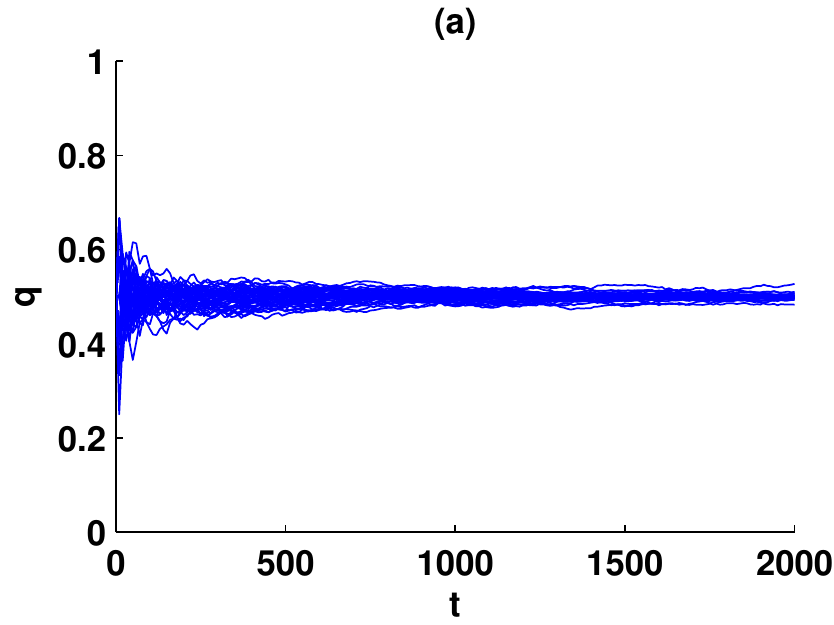}
\includegraphics[scale = 0.6]{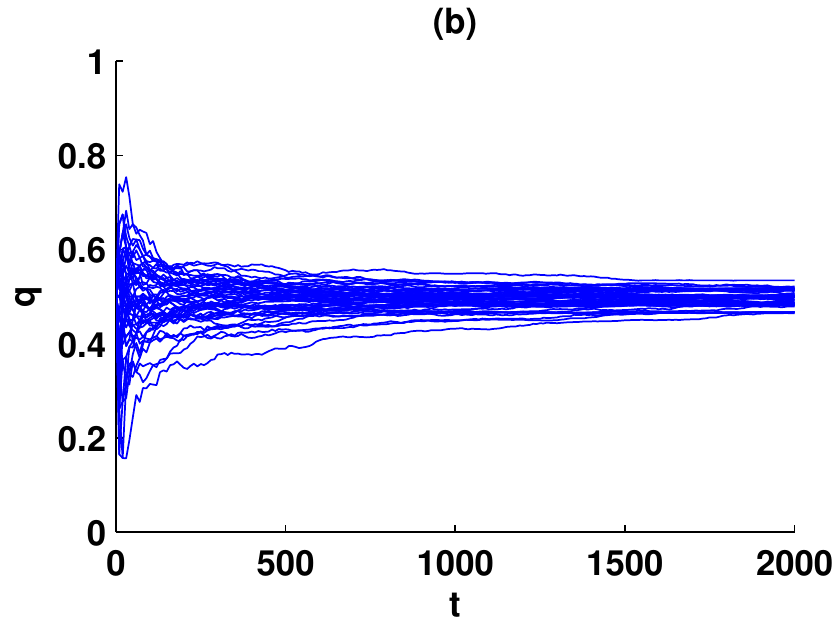}
\includegraphics[scale = 0.6]{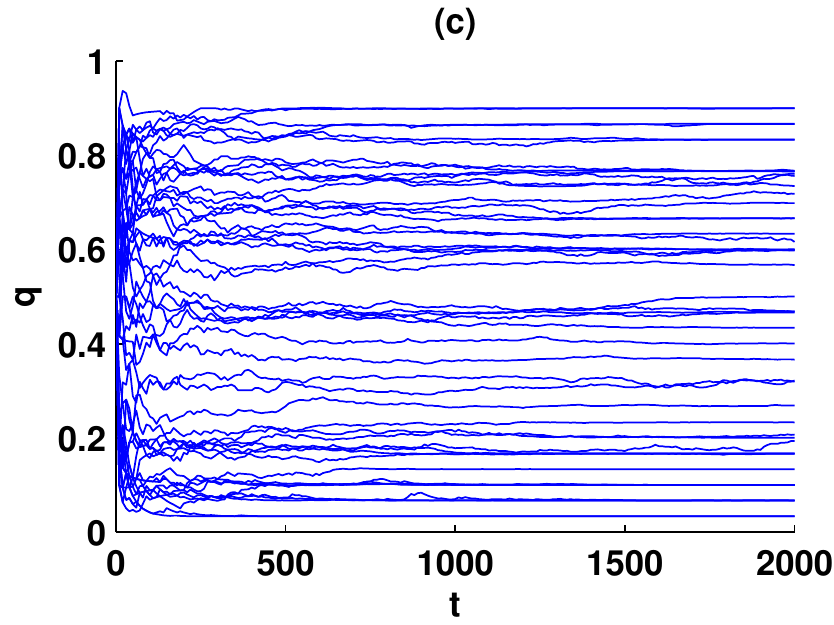}
\includegraphics[scale=0.6]{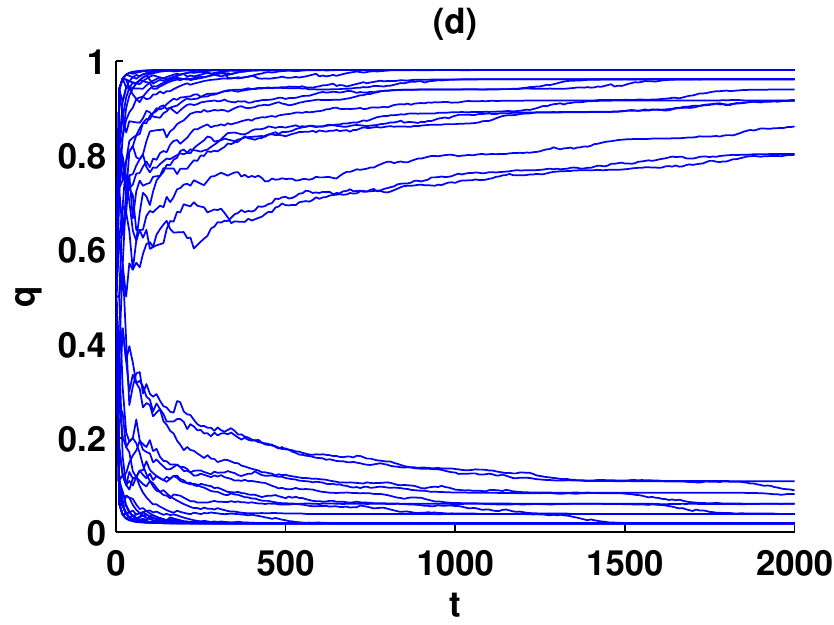}
\includegraphics[scale=0.6]{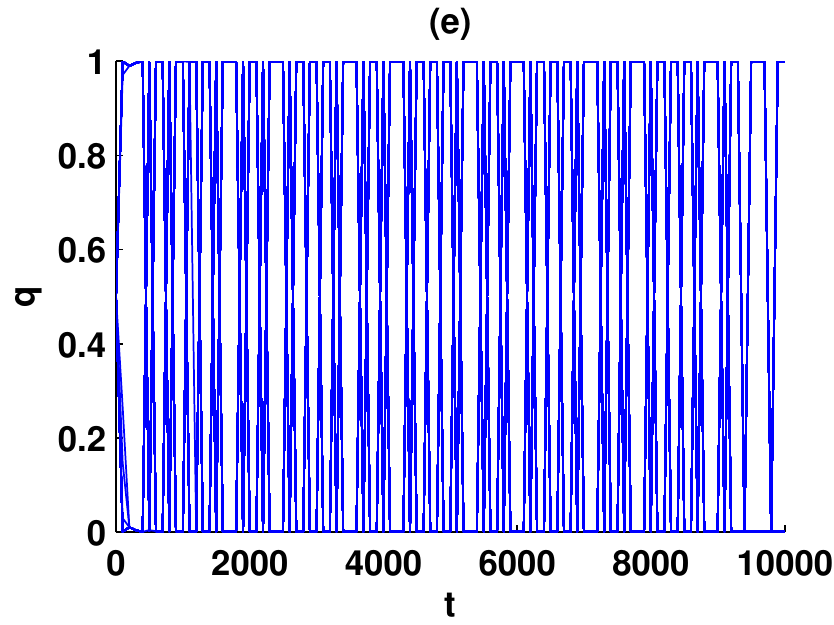}
\caption{This figure shows how the objective bias $q$ of the coin varies with time $t$
for the different reality maps illustrated in Figure~\ref{realityMaps}.
All the players have the same initial wealth.
The first player has fixed strategy
$p(\text{heads}) = 1/30$, $p(\text{tails}) = 29/30$, the second player has
fixed strategy $p(\text{heads}) = 2/30$, $p(\text{tails}) = 28/30$, etc. 
Figures (a)--(e) plot the objective bias $q$ of the coin as a
function of time using different random number seeds to generate the sequence of coin tosses.  Figure~(a) shows the self-defeating case, $q(p) = 1-p$, corresponding
to the reality map of Figure~\ref{realityMaps}(a): here,
the objective bias converges rapidly to the fixed point
of the map, $p=q=1/2$.  Players with this strategy eventually
acquire all the wealth.  Figure~(b) shows the weakly self-reinforcing
case $\alpha = 0.5$, corresponding to the reality map of
Figure~\ref{realityMaps}(b): in this case, the objective bias also converges
to the fixed point $p=q=1/2$, albeit more slowly than in
the self-defeating case of (a).  Figure~(c) shows the `purely subjective' case of Figure~\ref{realityMaps}(c), $q(p) = p$:
here, all points are fixed points.  In the purely subjective case,
the map eventually converges to one player's strategy ---
the strategy to which it converges is determined randomly
from the initial tosses of the coin.  Figure~(d) shows the strongly self-reinforcing case
$\alpha = 1.5$ of Figure~\ref{realityMaps}(d): here the objective bias gradually
moves either towards the fixed point $q=0$ or to the fixed point
$q=1$.  Finally, figure~(e) shows the behavior
under the multi-modal reality map $q(p) = 3p~\mod~1$ shown in Figure~\ref{realityMaps}(e).  The bias $q$
oscillates wildly and stochastically back and forth
between $0$ and $1$.  Although it is not obvious from the figure,
over time, the wealth becomes concentrated on strategies
$s=1/3$, $s=2/3$.}
\label{biasDynamics}
\end{center}
\end{figure}

In this section we study the dynamics of the objective bias of the
coin, which is the tangible reflection of ``reality'' in the game.
This also allows us to get an overview of the behavior for different
 reality maps $q(p)$.  We use $N= 29$ agents each playing one of the 29 equally spaced strategies on $(0,1)$: $1/30, 2/30, \dots, 29/30$, and begin by giving
them all equal wealth.  We then play the game repeatedly and plot
the bias of the coin $q^{(t)}$ as a function of time.  This is done
several times using different random number seeds to get a feeling for the variability vs.\ consistency
of the behavior of different reality maps, as shown in
Figure~\ref{biasDynamics}.

For the purely objective case $q(p) =
1/2$, the result is trivial since $q$ doesn't change.  For the self-defeating case,  $q(p) =
1 - p$, the results become more interesting, as shown in
(a). Initially the bias of the coin varies considerably, with a range that is
 generally about $0.3$ -- $0.7$, but it eventually settles into a fixed
point at $q = 1/2$.  For this case the bias tends to
oscillate back and forth as it approaches its equilibrium value. Suppose, for example, that the
first coin toss yields heads; after this toss, players who bet more
on heads possess a majority of the wealth.  At the second toss,
because of the self-defeating nature of the map, the coin is biased
towards tails.  As a result, wealth tends to shift back and forth
between heads and tails players before finally accruing to players
who play the `sensible' unbiased strategy.

We then move to the weakly self-reinforcing case using
equation~(\eqref{qalpha}) with $\alpha = 1/2$, as shown in (b).  The
behavior is similar to the previous case, except that the
 fluctuations of $q^{(t)}$ are now larger.  At the end of $2000$ rounds
of the game, the bias is much less converged on $q = 1/2$.  The bias
 is also strongly autocorrelated in time --- if the bias is high at a given
time, it tends to remain high at subsequent times.  (This was
already true for the self-defeating case, but is more pronounced
here.) Although this is not obvious from this figure, after a
sufficiently long period of time all trajectories eventually
converge to $q = 1/2$.

Next we study the purely subjective case, $q(p) = p$, as shown in
(c).  In this case the bias fluctuates wildly in the early rounds of
the game, but it eventually converges to one of the strategies
$s_i$, corresponding to the player who eventually ends up with all
the wealth.

As we increase $\alpha > 1$, as shown in (d), the instability
becomes even more pronounced.  The bias initially fluctuates near $q
 = 1/2$, but it rapidly diverges to fixed points either at $q
= 0$ or $q = 1$.  Which of the two fixed points is chosen depends on
the random values that emerge in the first few flips of the coin;
initially the coin is roughly fair, but as soon as a bias begins to
develop, it is rapidly reinforced and it locks in.  The extreme case
occurs when $q(p)$ is a step function, $q(p) = 0$ for $0 \le p <
 1/2$, $q(1/2)=1/2$ and $q(p) = 1$ for $1/2 < p \le 1$.  In this case the first
coin flip determines the future dynamics entirely: if the first coin
flip is heads, then players who favor heads gain wealth relative to
those who favor tails, and the coin forever after yields heads,
until all the wealth is concentrated with the player that bets most
heavily on heads.  (And vice versa for tails.)

Finally, in (e) we show an example of the bias dynamics for the
multi-modal map $q(p) = 3p \mod 1$.  In this case the bias
oscillates between $q = 0$ and $q = 1$, with a variable period that
is the order of a few hundred iterations.  We explain this behavior
at the end of the next section.

\section{Wealth dynamics}

How do the wealths of individual players evolve as a function of
time?  The purely objective case $q = \textit{const}$ with fixed
strategies and using a bookmaker instead of pari-mutuel betting was studied by \cite{Kelly56}, \cite{CoverThomas:book}, and \cite{Blume93}.  Assuming
all the strategies are distinct, they show that the agent with the
strategy closest to $q$ asymptotically accumulates all the wealth.
Here ``closeness'' is defined in terms of the
Kullback-Leibler distance\footnote{
The Kullback-Leibler distance $D_{KL}$ between two discrete probability distributions $P_i$ and $Q_i$ is defined as $D_{KL} (P \| Q) = \sum_i P_i \log P_i/Q_i)$.}
 between the strategy vector $s_{il}$ and the true
 probability vector of objective outcomes $q_l$.

For all reality maps $q(p)$ that we have studied we find
that one player asymptotically accumulates nearly all the wealth.
As a particular player becomes more wealthy, it becomes less and
less likely that another player will ever overtake this player.
 This concentration of wealth in the
hands of a single player is the fundamental fact driving the
convergence of the objective bias dynamics to a fixed point, as
observed in the previous section.  The reason is simple:  once one
player has all the wealth, this player completely determines the
odds, and since her strategy is fixed, she always places the same
bets.

\subsection{Purely subjective case}

\begin{figure}[htbp]
\begin{center}
\includegraphics[width=0.45\textwidth]{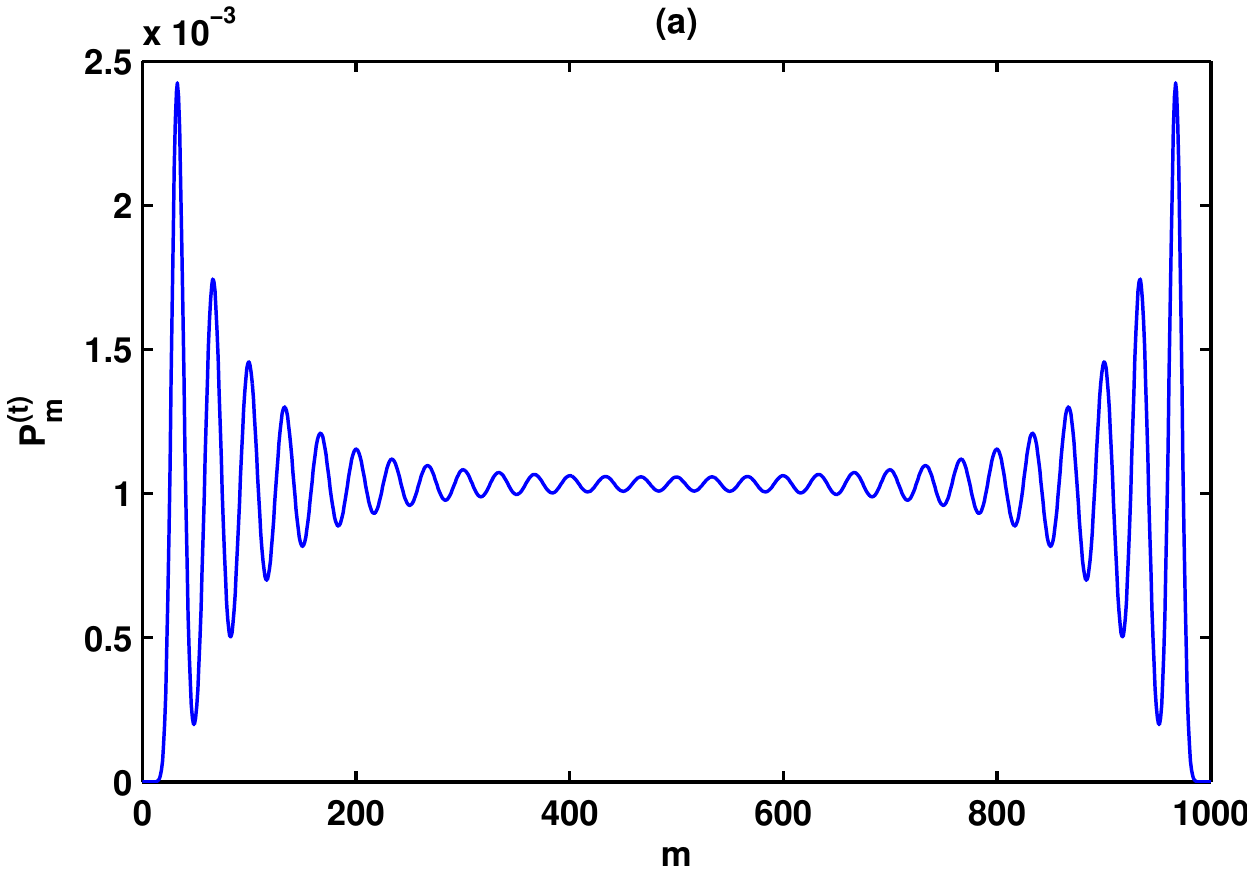}
\includegraphics[width=0.45\textwidth]{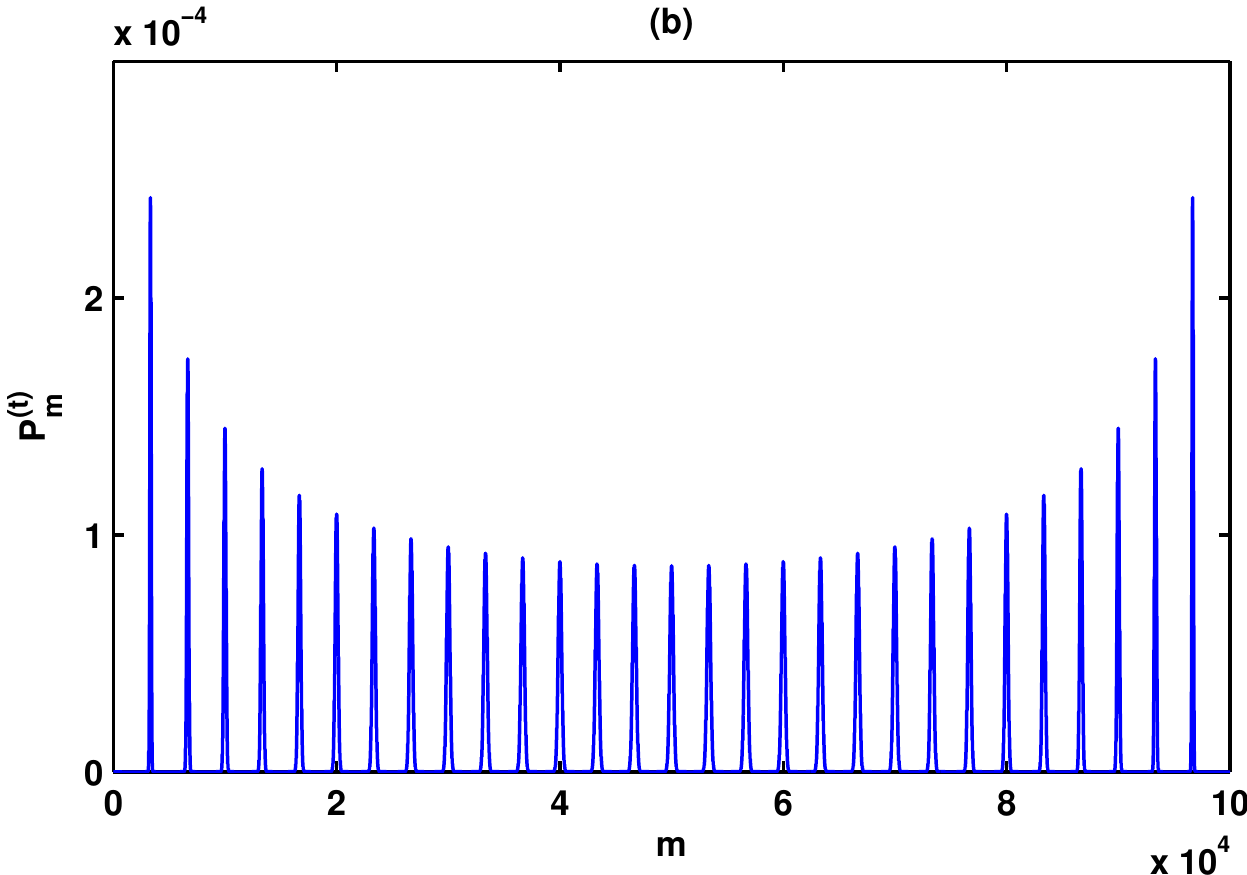}
\caption{In the purely subjective case, $q(p) = p$, all fixed betting strategies (e.g., betting $1/3$ on heads and $2/3$ on tails) have an equal chance of eventually dominating the game. As before, there are $N = 29$ players with the same initial wealth, each playing a different fixed strategy. This figure shows the probability $P_m^{(t)}$ as a function of $m$, where $P_m^{(t)}$ is the probability that heads occurs $m$ times in $t$ rounds of the game with $q(p)=p$ for (a) $ t=10^3$ and (b) $t=10^5$.  After $10^3$ tosses in (a), the probabilities of heads occurring $m$ times in $t$ rounds have begun to show pronounced peaks around points $m=t\frac{k}{N + 1}$, $k=1,...,29$, which correspond to individual players' strategies. In (b), after $10^5$ tosses, the probabilities of ending close to one of the players' strategies are very pronounced and clearly separated; also note that, although some peaks are higher than others, the total weight (the integral) of each peak is the same.  This means that each fixed betting strategy has the same probability of eventually dominating the game.}
\label{binomialWealth}
\end{center}
\end{figure}

It is possible to compute the distribution of wealth after $t$ steps
in closed form for the purely subjective case, $q(p) = p$.  The
probability that heads occurs $m$ times in $t$ steps is a sum of
binomial distributions, weighted by the initial wealths $w_i^{(0)}$
of the players,
\begin{equation}
P_m^{(t)} = \sum_{j=1}^N
\w{j}{0}\left[{\binom{t}{m}}s_j^m(1-s_j)^{t-m}\right] \,,
\label{binomial}
\end{equation}
and the corresponding wealth of player $i$ is
\begin{equation}
\w{i}{t}= \D \frac{s_i^m(1-s_i)^{t-m}\w{i}{0}}{\sum_j
s_j^m(1-s_j)^{t-m}\w{j}{0}}\,.
\label{binomialwealth}
\end{equation}
When the initial wealths are evenly distributed among the players,
no player has an advantage over any other.  However, as soon as the
first coin toss happens, the distribution of wealth becomes uneven.
Wealthier players have an advantage because they have a bigger
influence on the odds, so the coin tends to acquire a bias
corresponding to the strategies of the dominant (i.e.\ initially
lucky) players.  Figure~\ref{binomialWealth} shows the probability
$P_m^{(t)}$ as a function of $m$ for $ t=10^3$ and $t=10^5$, with $N=29$.
After $10^3$ steps the
binomial distributions are still strongly overlapping, and there is
still a reasonable chance to overtake the winning strategy.  After
$10^5$ steps, however, the bias of the coin has locked onto an
existing strategy $s_i$, due to the fact that this strategy has
almost all the wealth.  Once this happens, the probability that this
will ever change is extremely low.

\subsection{Multi-modal reality map}

We now explain the peculiar bias dynamics observed in Figure~\ref{biasDynamics}(e) for
the multi-modal map $q(p) = 3p \mod 1$ of
Figure~\ref{realityMaps}(e), in which the bias of the coin
oscillates wildly between $0$ and $1$.  Through the
passage of time the wealth becomes concentrated on strategies near
either $s = 1/3$ or $s = 2/3$, corresponding to the
discontinuities of $q(p)$.  Suppose, for example, that $p = \sum_i
w_i s_i$ is slightly greater than $1/3$, where the $q(p)$ map is
close to zero.  This causes a transfer of wealth toward strategies
with smaller values of $s$ until $p = \sum_i w_i s_i < 1/3$.  At
this point the bias of the coin flips because $q(p)$ is now close to
one and the transfer of wealth reverses to favor strategies with
higher values of $s$.  Due to fluctuations in the outcomes of the
coin tosses this oscillation is not completely regular.   It continues
indefinitely, even though with the passage of time wealth becomes
concentrated more and more tightly around $s = 1/3$.  A
similar process occurs if the first coin tosses cause convergence around $s = 2/3$.  We discuss the initial convergence around $s =1/3$ or $s = 2/3$ in more detail in Section~\ref{sec:rational-players}.

\section{Analytic analysis of approach to equilibrium}
\label{analyticSection}

So far, our discussion of the `reality game' has been heuristic
and based on simulations.  In this section we develop an
analytic description of the game when players play fixed
strategies.  In this simple case the behavior of the
game in general, and the approach to equilibrium\footnote{
Asymptotically as $t \to \infty$ the wealth always becomes concentrated on a single strategy.   This is what we mean by ``equilibrium". As described in subsequent sections, once the system reaches equilibrium it is no longer possible for an optimal player to make profits, i.e. the game is perfectly efficient.}
in particular,
admit a closed-form analytic solution.

This section
establishes the following features of the reality game
with fixed strategies:

\begin{itemize}
\item
Stable equilibria of
the game correspond to fixed points $\tilde{q} = q(\tilde p)= \tilde p$ of
the reality map, with $\mu = q'(\tilde p)  < 1$.
\item
A central-limit theorem argument shows that the distribution
of wealth among the different strategies approaches a
Gaussian distribution after many plays of the game.
\item
The approach to the fixed point is governed by a power law.
\end{itemize}

The method that will be used is as follows.  First, we
establish that the wealth distribution tends
to a Gaussian.	Next, we derive a hierarchy of equations for
the approach of the moments of this distribution to the
fixed points; in this hierarchy, the rate of approach of the $k$-th
moment will depend on the value of the $(k+1)$-th moment.
The Gaussian character of the wealth distribution allows us
to truncate the hiearchy of equations to obtain
coupled ordinary differential equations for the mean
and variance.  Finally, we solve these equations to
show that the approach to equilibrium is governed by a power law.
In the process of solution we find that
fixed points $\tilde p = 0,1$ behave differently from
fixed points $\tilde p \neq 0,1$.

\subsection{Wealth updating}

Assume a continuum of strategies
$\{s\}$, $0\leq s \leq 1$.	Let $w(s)$ be the distribution
of wealth over the strategies with $\int w(s) ds = 1$.
The total amount bet on heads is
$p = \int s w(s) ds $.
The updating rule is as follows.  Let $w_t(s)$ be the wealth
distribution at time $t$.  If the coin comes down heads, from
equation~(\eqref{expectwealth}) the new distribution is
\begin{equation}
w_{t+1,h}(s) = s w_t(s)/p_t.
\label{w1h}
\end{equation}
Similarly, if it comes down tails, the new distribution is
\begin{equation}
w_{t+1,t}(s) = (1-s) w_t(s)/(1-p_t).
\label{w1t}
\end{equation}

\subsection{Wealth distribution approaches a Gaussian}

It is simple to show that the wealth distribution $w_t(s)$ approaches
a Gaussian after many trials.
After many trials the distribution is proportional to
$s^m (1-s)^n w_0(s)$, where $m$ is the number of heads and $n$
is the number of tails.	 In the limit as $m$ and $n$ become
large, this is an ever narrower binomial  distribution
multiplied by $w_0(s)$: as long as the derivatives of
$w_0(s)$ are finite, the only part of $w_0(s)$ that is important
is its value at the peak.  Thus the overall
distribution tends to a binomial distribution, which asymptotically becomes a Gaussian distribution scaled by the value of $w_0(s)$ at the peak.
For $m<<n$, we can also use the Poissonian approximation to
the binomial distribution.  For the analysis below, however,
both Poissonian and Gaussian approximations give the same
dynamics: accordingly, we will use the Gaussian approximation below.

Suppose that there have been $t$ trials, so that $m+n =t$.
The average value of the approximately Gaussian distribution $w_t(s)
\propto s^m (1-s)^n w_0(s)$ is then $p = m/t$, and the
variance of the Gaussian is $D = p(1-p)/t$.  This formula
for the variance will prove useful for determining the behavior 
of the game as it approaches equilibrium.

\subsection{Dynamics of average bet}

Now derive a formula for the change in the amount
bet on heads over time.	 At time $t$, the amount
bet on heads is the mean of the wealth distribution,
$p_t = \int s w_t(s) ds $.  Define the variance of the
distribution to be
$D_t = \int s^2 w_t(s) ds - p_t^2$.  If the coin comes
down heads, then using equation~(\ref{w1h}) the new value of $p$ is
\begin{equation}
p_{t+1,h} = \int s w_{t+1,h}(s) ds =  \int \frac{s^2 w_t(s)}{p_t} ds  =
 p_t + \frac{D_t}{p_t}.
\label{p1h}
\end{equation}
That is, if the coin comes down heads, the
value of $p_t$ is displaced by $ D_t/p_t$
in the heads direction.
Equation~\eqref{p1h} does not rely on the 
Gaussian approximation.  Inserting the formula 
$D= p(1-p)/t$, we obtain a formula
for the displacement of $p$ in the Gaussian approximation:
\begin{equation}
p_{t+1,h} \eqGaussian p_t + \frac{(1-p_t)}{t},
\label{p1hG}
\end{equation}
where $\eqGaussian$ indicates equality in the Gaussian approximation.

In the same way, we obtain formulae for the behavior
of the system if the coin comes down tails.  If the coin comes down tails,
the new value of $p$ is
\begin{equation}
p_{t+1,t} = \int \frac{s(1-s) w_t(s)}{1- p_t} ds =
 p_t - \frac{D_t}{1- p_t}.
\label{p1t}
\end{equation}
That is, if the coin comes down tails, the
value of $p$ is displaced by $ D_t/(1-p_t)$
in the tails direction.  In the Gaussian approximation,
this becomes
\begin{equation}
p_{t+1,t} \eqGaussian p_t - \frac{p_t}{t}.
\label{p1tG}  
\end{equation}

On the $(t+1)$-th toss, the coin comes up heads with probability
$q_t$ and tails with probability $1-q_t$.
Define $\Delta p_t = p_{t+1} - p_t$.  Letting $E$ represent the expectation over heads and tails,
from equations~(\eqref{p1h}) and (\eqref{p1t}) the average
change in $p$ is then
\begin{equation}
\E[\Delta p_t] = q_t p_{t+1,h} + (1-q_t) p_{t+1,t} - p_t
= q_t \frac{D_t}{p_t} - (1-q_t) \frac{D_t}{1-p_t}
= (q_t - p_t) \frac{D_t}{p_t(1-p_t)},
\label{deltap0}
\end{equation}
and for Gaussian approximation, the corresponding expression
is obtained from equations~(\eqref{p1hG}) and (\eqref{p1tG}):
\begin{equation}
\E[\Delta p_t] \eqGaussian q_t\frac{(1-p_t)}{t} - (1-q_t)\frac{p_t}{t}
= \frac{q_t - p_t}{t}.
\label{deltap0G}
\end{equation}
In the regime where the Gaussian approximation is
valid, the system takes a random walk with a step
size that decreases as $1/t$.

\subsection{Approach to equilibrium}

Now employ the functional dependence of $q$ on $p$.
In the vicinity of the fixed point $\tilde q = q(\tilde p) = \tilde p$,
we can use a linear approximation of $q(p)$ and write
\begin{equation}
q_t  \approx \tilde q + \mu ( p_t - \tilde q), 
\label{linapprox}
\end{equation}
where $\mu = q'(\tilde p)$ is the slope at the fixed point. 
Substituting the linear approximation of equation~(\eqref{linapprox})
into the expression for the average change
in $p$, equations~(\eqref{deltap0}) and (\eqref{deltap0G}),
we obtain (after some algebra)
\begin{equation}
\E[\Delta p_t] =  \frac{D_t (\mu-1) ( p_t - \tilde q )}{p_t (1 - p_t)}
\eqGaussian (\mu-1) \frac{p_t - \tilde q}{t}.
\label{deltap1}
\end{equation}
Defining the distance to the fixed point to be
$y = p - \tilde q$, and letting $\Delta y_t =  y_{t+1} - y_t$,
we now have an equation for
the average change in this quantity:
\begin{equation}
\E[\Delta y ]  = \frac{D(\mu-1)}{p (1 - p)} y
= \frac{D(\mu-1)}{(\tilde q + y )  (1 - \tilde q - y)} y 
\eqGaussian (\mu-1) \frac{y}{t}.
\label{dydt0}
\end{equation}

From equation~(\eqref{dydt0}), we see right away that the condition for
converging to a fixed point is that the slope at the
fixed point, $\mu = q'(\tilde p)$, is less than one.
This was also seen numerically in Section~\ref{sec:dynamics-of-bias}.

In the Gaussian limit, equation~(\eqref{dydt0}) shows
that the expected change in $y$ from one
step to another is linear in $y$.  As a result,
it can be shown by explicit calculation that the
change in $\E[ \Delta y_t ]$ from time $t$ to $t+1$
is equal to $ \Delta \E[ y_t] $,  
where $\E[y_t]$ is the expected value of the distance
from the equilibrium point after $t$ time steps.
In other words, rather than tracking the entire stochastic
process and taking its average value at a time, we can
simply construct a differential equation for the
average value $\E[ y_t ]$.  In particular,
averaging again in equation~(\eqref{dydt0}) then yields an equation
for $\E[y_t]$:
\begin{equation}
\Delta \E[y_t] = (\mu-1) \frac{\E[y_t]}{t}.
\label{dydt} 
\end{equation}
In the following, for the sake of compactness,
define $\bar y_t = \E[y_t]$.
Writing equation~(\eqref{dydt}) in continuous time gives
a differential equation for $\bar y$:
\begin{equation}
\frac{d\bar y}{dt} = (\mu-1) \frac{\bar y}{t}.
\end{equation}
In other words, in the Gaussian regime, 
the approach to
equilibrium obeys a power law on average:
\begin{equation}
\bar y \propto  t^{\mu -1}.
\label{yt}
\end{equation}
The rate of approach to equilibrium depends on the slope
$\mu$ of the reality map $q(p)$ at the fixed point.

\subsection{Fluctuations about mean}

It is important to keep in mind that equation~(\eqref{yt}) is 
an equation for the average value of $y$: for any given
set of coin flips, there will be fluctuations about
that average.  These fluctuations are governed by
the variance $\Delta_t$ of the stochastic process.  
As noted above, $y$ takes a biased random walk with step size that
goes as $O(1/t)$ for late times.  These late time fluctuations
can be shown to have an average magnitude
$\sqrt{p(1-p)/t}$, the same as the fluctuations in the
wealth distribution at time $t$. 

This estimate of the late time fluctuations does not include
the propagation forward in time of earlier fluctuations.
Early on in the random walk, the step sizes are larger: later on,
they are smaller.  The early fluctuations propagate forward in
time like $t^{(\mu-1)}$.  As we'll now show,
when one adds in all fluctuations that occur
at all points in time, these propagating earlier fluctuations
combine to give an average fluctuation size of $\Delta_t = t^{(\mu-1)/2}$.
More precisely, we have the following accounting,
using the Gaussian approximation, and the biased random
walk of section (5.3).

A fluctuation of size $f$ at time $t_1$ propagates over time as 
$f \times  (t/t_1)^{(\mu-1)}$.  Let's use this feature to 
propagate forward in time the effect of all fluctuations. 
At each time step, the magnitude of the fluctuation is 
on the order of the step size.
The first step is of size $O(1)$, the second step is of size $O(1/2)$,
the third step is of size $O(1/3)$, etc., with the $k$'th step
being of size $O(1/k)$.  So our overall fluctuations are of the
following order:
\begin{equation}
\Delta_t = \pm 1 \times \big({t\over 1}\big) ^{\mu-1}
\pm {1\over 2} \times \big({t\over 2}\big)^{\mu-1}
\pm {1\over 3} \times  \big({t\over 3}\big)^{\mu-1}
\pm  \ldots \pm {1\over t} \times \big({t\over t}\big)^{\mu-1}
= t^{\mu-1} \bigg( \pm {1 \over 1^\mu} \pm {1 \over 2^\mu}
\pm {1 \over 3^\mu} \pm \ldots \pm {1 \over t^\mu} \bigg)
\label{Deltat}
\end{equation}

That is, the $k$'th term in this sum goes as $t^{\mu-1}$ times plus
or minus $1/k^\mu$.   But the typical magnitude of the sum
of plus or minus $1/k^\mu$ from $1$ to $t$ is on the order
of $t^{(1-\mu)/2}$.  (This estimated magnitude relies on
a Gaussian approximation: the accuracy of this estimate
will decline as $\mu \rightarrow 1$ for $\tilde q \neq 0,1$,
and as $\mu\rightarrow 0$ for $\tilde q = 0,1$.  For future work,
it would be useful to obtain analytic solutions for the scaling of the
fluctuations in these two limits.)
As a result, the estimated magnitude of
the early time fluctuations, propagated forward and summed, goes as
$\Delta_t = O(t^{(\mu-1)/2})$.  
If $\mu \geq 0$, the total fluctuations
are dominated by early time fluctuations, propagated forward in time.
If $\mu < 0$, then it is only the late-time fluctuations 
that are important, and these go as $t^{-1/2}$, as
mentioned above.  Since the average distance from a fixed point
shrinks as $y_t \propto t^{\mu-1}$, and as fluctuations
go either as $\Delta_t \propto t^{(\mu-1)/2}$ ($\mu \geq 0$),
or $\Delta_t \propto t^{-1/2}$ ($\mu < 0$), the actual
distance from a fixed point at any time is dominated by
fluctuations about the mean value of $y_t$. 

\subsection{Summary}

To summarize this section, the reality game with
fixed strategies exhibits non-trivial but still
analytically tractable behavior.  Equilibria correspond to
fixed points of the reality map with slope less than one,
and the mean distance from those equilibria shrink as a
power law in time with exponent $\mu-1$, where $\mu$ is the
slope of the reality map at the fixed point.  The actual distance
from equilibrium at any time is dominated
by fluctuations about that mean value.  

When players possess non-fixed strategies, the behavior
of the game becomes more complicated.  Equilibria still
exist, however.	 In the next section, we introduce
rational players who change their strategies over
time.

\section{Rational players and the long-term equilibria}
\label{sec:rational-players}

In this section we will show that by introducing an appropriate notion of rational players it is possible to predict the set of possible asymptotically dominant strategies, i.e.\ the long-term equilibria.  To do this we imagine a player who is rational in the strong sense that she not only understands the structure of the game, but also knows all the strategies and wealths of all the other players.   The rational player is able to use this information to optimize her play.  The players placing bets with fixed strategies are obviously sub-rational in any sense of the word.  Such models with a mixture of rational and sub-rational players are often called \textit{noise trader} models in the finance literature (\cite{Shleifer00}).

Use of a rational player requires the choice of a utility function.  As originally shown by \cite{Kelly56}, logarithmic utility is the unique choice that optimizes the asymptotic growth rate of wealth.\footnote{
Although Kelly's result is applicable only in case of
infinitely long games, it's possible to show
(\cite{Cherkashin04}) that it also holds for finite games
provided that their duration is not known to the players.
So in this case too, the logarithmic utility happens to be
a rational choice for the players, and the players should
maximize their asymptotic growth rate of wealth.}
If one player's  asymptotic growth rate is higher than that of all the others, that player will end up with all the money.  This dictates that  for our purpose, which is to predict the asymptotically dominant strategies, we want to consider rational players who maximize log-returns.\footnote{
If rational players maximize objective functions other than logarithmic utility it is possible to produce examples where non-rational players do a better job of maximizing logarithmic utility, and are therefore more likely to survive (\cite{Blume93}, \cite{Sandroni00}).}
The choice of other utility functions can be disastrous.  For example, maximizing simple returns often dictates betting everything on a single outcome.  Doing this an infinite number of times guarantees that eventually some other outcome will be selected and the player will lose all her wealth, which is obviously a poor strategy if one's goal is to maximize the probability of long-term survival.


\subsection{Derivation of optimal strategy}

For the case $q = \textit{const}$, as shown in \cite{CoverThomas:book}, the asymptotic log-returns can be maximized by simply repeatedly maximizing the expected log-return on the next step.
\begin{eqnarray*}
\E\r{i}{t+1} & = & \E \log \frac{\w{i}{t+1}}{\w{i}{t}} \\
& = & \E\log\frac{\pi_i}{\w{i}{t}} = \sum_l q_l \log \frac{s_{il}}{p_l}.
\label{logreturn}
\end{eqnarray*}
The reason this works is because when $q = \textit{const}$, the actions of the players do not affect the odds of the coin, and the player does not have to take the future bias of the coin into account in optimizing her strategy. For a more general reality map $q(p)$ that incorporates subjective feedback, this is no longer the case.  It is easy to produce examples in which maximizing the log-return two steps ahead produces different results than successively maximizing the log-return a single step ahead.  Nonetheless, we find that one step maximization correctly predicts the long-term equilibria.  We return to discuss this further later in this section.

Under the assumption of one step maximization, we define a (myopic) optimal player as one that maximizes
the expected log-return on the next step. The expected log-return for one step can be written
\begin{equation}
r(s) = q\log\frac{s}{p} + (1-q)\log\frac{(1-s)}{(1-p)},
\label{maxlogreturn}
\end{equation}
where as before $q = q_1$, and $p = p_1$, i.e. these are the values for heads, and $s$ is the bet of the rational player on heads.  
Recalling that for a discrete strategy set $p = \sum_i s_i w_i$, or for a continuous set $p = \int s w(s) ds$, the first derivative $dr/ds$ is
\begin{multline}
\frac{dr}{ds} = q'w\bigg[\log s - \log p - \log(1-s) + \log(1-p)\bigg] \\
 + q\left(\frac{1}{s} - \frac{w}{p}\right) + (1-q)\left(-\frac{1}{1-s} + \frac{w}{1-p}\right),
\label{drds}
\end{multline}
where $q' = dq/dp$.
%

\subsection{Increasing returns}

\begin{figure}[htbp]
\begin{center}
\includegraphics[width=0.6\textwidth]{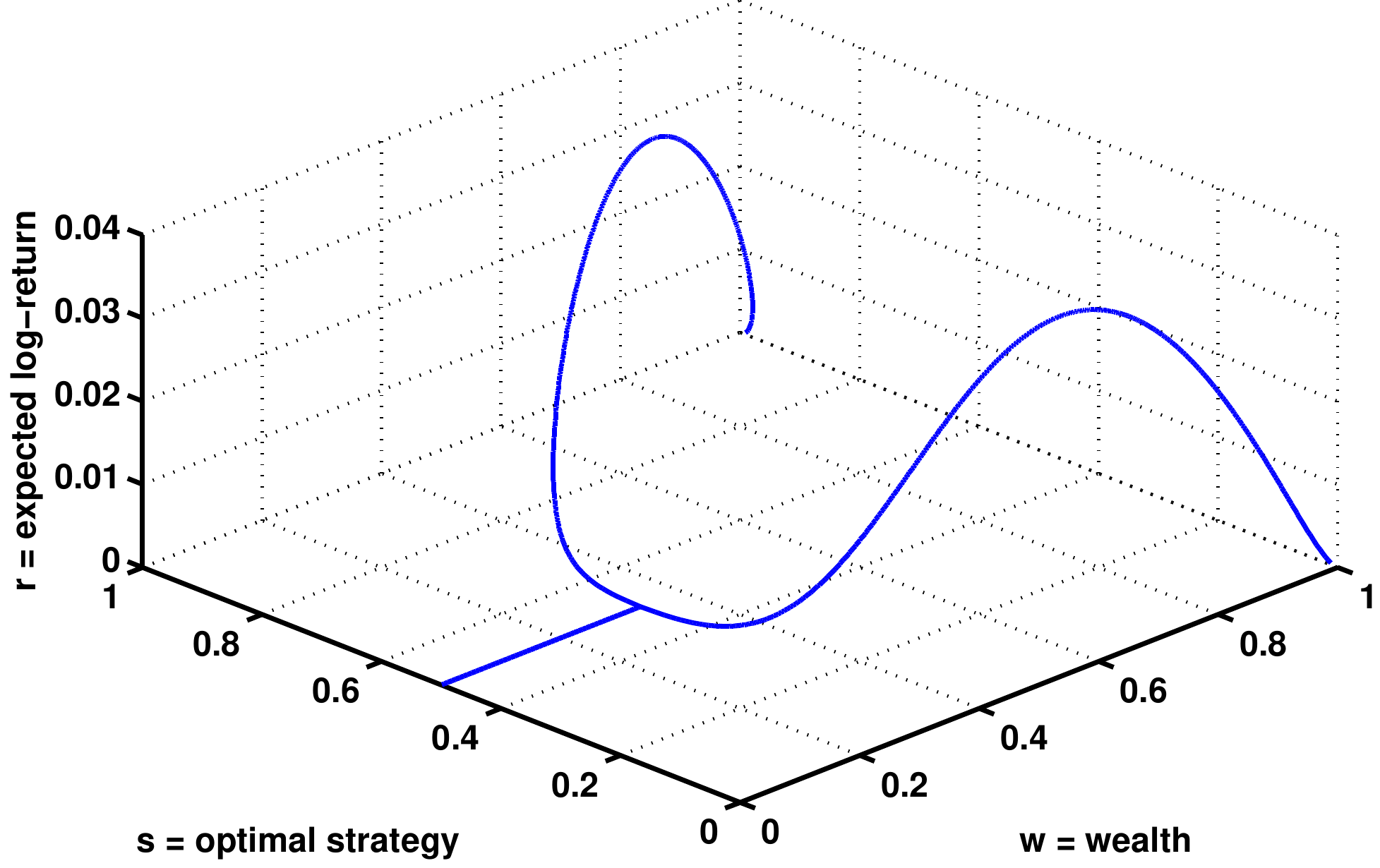}
\caption{When the game is self-reinforcing the rational strategy is sometimes able to generate increasing returns, i.e.\ its returns may increase as it gains in wealth. This figure shows the expected log-return $r(s)$ for a rational player with wealth $w$ using a myopic optimal strategy $s$ against a second player with wealth $1 - w$ using fixed strategy $s =1/2$. The reality map is strongly self-reinforcing: $q(p) = q_\alpha(p)$ for $\alpha = 2$.  If the optimal player's wealth is less than $1/3$ her best bet is the equilibrium strategy at $s = 1/2$, which only allows her to break even with the fixed strategy player, i.e. her log returns are zero.  When her wealth is more than this, however, the fixed point becomes unstable and she can use her ability to influence the objective probability to drive it away from $q=1/2$ and towards one of the two other fixed points $q=0$ or $q=1$, thereby making profits at the expense of the fixed strategy player.}
\label{nashEq}
\end{center}
\end{figure}

In general the optimal strategy depends on the reality map $q(p)$ and also depends on the wealth of the other players.  The optimal strategy can be dynamic, shifting with each coin
toss as the wealth of the players changes.
This is illustrated in
Figure~\ref{nashEq}.  Here we show the expected log-return for a
rational player with wealth $w$ playing against a second player with wealth $1 - w$, where the second player uses a fixed strategy with $s_2 = 1/2$.  We compute the strategy of the rational player by numerically solving $dr/ds = 0$, where $dr/ds$ is from equation~(\eqref{drds}).
When the rational player's wealth is low, $s = 1/2$ is the optimal
strategy, and she can only break even with the fixed player.  As the rational player
gains in wealth, two strategies on either side of $s = 1/2$ become
superior, and the rational player makes positive log-returns.  As her wealth $w$ increases, the optimal strategies become more and more separated from $s =
1/2$, and in the limit as $w \to 1$ the optimal strategy is either
$s = 0$ or $s = 1$.



\subsection{Equilibria}

The equilibria of the game can be simply computed by noting that that in equation~(\eqref{drds}) a sufficient condition for $dr/ds = 0$ is $s = q = p$, i.e.\ the strategy corresponding to the fixed point $q = p$.  This may or may not be stable, depending on the condition of the second derivative.  When it is stable, however, it corresponds to an attracting fixed point.
This will be stable when the second derivative satisfies
\begin{equation}
\frac{d^2 r}{ds^2} = \frac{1-w}{s(1-s)}\bigg[2wq'-(1+w)\bigg] < 0.
\label{d2rds2}
\end{equation}
When all the wealth is sufficiently closely concentrated around the equilibrium a rational player need only play the equilibrium strategy.
This can easily be seen by assuming that the rational player begins with a very small amount of wealth.
The approach to equilibrium is then dominated by the fixed-strategy
players, and the rational player simply plays the optimal
strategy $s^{(t)}=q^{(t)}$.  She does not begin to dominate
the overall wealth until the game has come close to the
fixed-strategy equilibrium.  

What happens once the rational player obtains most of
the wealth?  If this player started off with only
a small amount of wealth, then she does not dominate until
the game has approached very closely to a particular
fixed-strategy equilibrium.  But at this point the wealth of the fixed strategy
players is concentrated around the equilibrium strategy, and thus
is close to the strategy of the rational player.  Consequently, once the game
is close to a fixed-strategy equilibrium, the fact that
the rational player has acquired most of the wealth
makes little difference to the approach to equilibrium.

More precisely, let $y = p -\tilde q$ be the distance
from the equilibrium point, as before.  The optimum strategy
$p = q(\tilde p + y)$ adopted by the rational player is by definition a
local maximum of the expected log-return $r$, as discussed
above.	The expected log-return of a player at the actual
equilibrium point, compared with that of the rational player, is then $r + (d^2 r/ds^2) y^2$, where
$d^2 r/ ds^2$ is defined in equation~(\eqref{d2rds2}) above. 
That is, the deviation of the
log-return of the fixed-strategy player at the equilibrium
point from the log-return of the rational player
goes as second order in $y$, and can be neglected
in the regime where the analysis of the previous section
on fixed strategies holds true.	 Accordingly, even
in cases where the rational player acquires most
of the wealth, as long as this does not happen until
the game is already close to a fixed point,
we expect the power-law approach
to equilibrium to hold to a high degree of accuracy.

Note also that the multimodal map $q(p) = 3p \mod 1$ is interesting because all the
intersections with the identity are local minima for the expected
log-return.  Instead, the system is attracted to the discontinuities
of the map at $s = 1/3$ and $s = 2/3$.  It is as if one can think of
the discontinuities of the map as being connected (with infinite
slope), creating intersections with the identity that yield local
maxima of the log-return.  

\subsection{Single step vs.\ multi-step optimization}

Under some circumstances it is possible to decompose the optimization into a series of single step optimizations.  To see this, consider in more detail the decomposition of the $T$-step log-return into a series of one step returns.  Let $r_t = r(s_i^{(t)}, p_t)$ be the expected log-return for strategy $s_i$ on step $t$.  This depends on the strategy $s_i^{(t)}$ used by player $i$ and the amount $p_t$ wagered on heads by all the players on step $t$.  Let $S_i^{(t-1)} = (s_i^{(1)}, s_i^{(2)}, \ldots, s_i^{(t-1)})$ be the sequence of strategies used by player $i$ on all previous steps.  For a non-constant reality map $q(p)$ the returns depend on the wagers of the players, which in turn depend on their wealths, which depend among other things on the previous moves of player $i$.  To emphasize this we write $p_t = p(w_t(S^{(t-1)}))$.  We can write the log-return over $T$ steps as
\begin{equation}
R_T = \sum_{t=1}^T r(s^{(t)}, p(w_t(S^{(t-1)}))).
\label{RT}
\end{equation}
This makes it clear why in general finding a strategy that maximizes log-returns over $T$ steps is challenging:  to properly optimize it is necessary to take into account how the bet at $t=1$ influences the wealth of all other strategies at all 
future times.  In some cases, however, this difficult optimization
problem can be finessed.  Consider the case where player $i$ has
negligibly small wealth at time $t = 1$.  Because $i$'s wealth is negligible,
it has no influence on the wealth of the other strategies, and
$p_t$ is independent of $S^{(t-1)}$ for this player.  In this case, player $i$ can
optimize over a horizon of length $T$ by performing a series
of successive single step optimizations, even when $q(p)$ is not
constant.  At the beginning of the game, when there are many 
players, each of whose wealths is small, and to a reasonable approximation all players can
optimize their strategies over a horizon of length $T$ by performing
single step optimizations.

\section{Efficiency}
\label{sec:efficiency}

As the game is played the reallocation of wealth causes the
population of players to become more efficient in the sense that
there are poorer profit opportunities available for optimal players.
This is analogous to financial markets, where standard dogma asserts that wealth reallocation
due to profitable vs.\ unprofitable trading will result in a market that is efficient in the sense
that all excess profit-making potential has been removed.  One would like to be able to measure market inefficiency, and to study the transition from an inefficient to an efficient market.  In this section we introduce a quantitative measure of inefficiency, study how it converges in time, and show that it exhibits interesting scaling properties.  In particular, the degree of inefficiency converges as a power law, and converges more slowly when there is a higher level of subjective feedback.

\subsection{Definition of inefficiency}

We define the inefficiency of our game to be the log-return for a \textit{rational $\epsilon$ player}.  Like the rational player discussed in the previous section, this player knows the
strategies of all other players, and pursues an optimal strategy
that maximizes her expected log-returns.  In addition, this player has
infinitesimal wealth $\epsilon$, so that her actions have a
negligible effect on the outcome of the game.\footnote{
In the limit $t \to \infty$, a rational $\epsilon$ player might accumulate all
the wealth, in which case she would no longer be an $\epsilon$ player.  Thus we assume that for any given $t$, $\epsilon$ is sufficiently small so this player's wealth remains negligible.}
As described in the previous section, the assumption of infinitesimal wealth is convenient because it means that the player need only optimize one step ahead, and because it guarantees that the rational $\epsilon$ player does not affect the wealth dynamics of the game.  The log-return for the rational $\epsilon$ player defines the inefficiency at every time step.

In the purely objective setting where $q = \textit{const}$, the approach to
efficiency is guaranteed by the fact that the wealth dynamics are
formally equivalent to Bayesian updating, implying all the wealth
converges on the correct hypothesis about the bias of the coin. For
more general settings this is no longer obvious, as there is no
longer such a thing as an objectively correct hypothesis.   Nonetheless, we always find the system converges to an equilibrium where all the wealth is concentrated on a single point.  At equilibrium there are no profit opportunities for the rational $\epsilon$ player, and thus the market is perfectly efficient.

\subsection{Numerical study of inefficiency vs. time}

\begin{figure}[htbp]
\begin{center}
\includegraphics[width=0.8\textwidth]{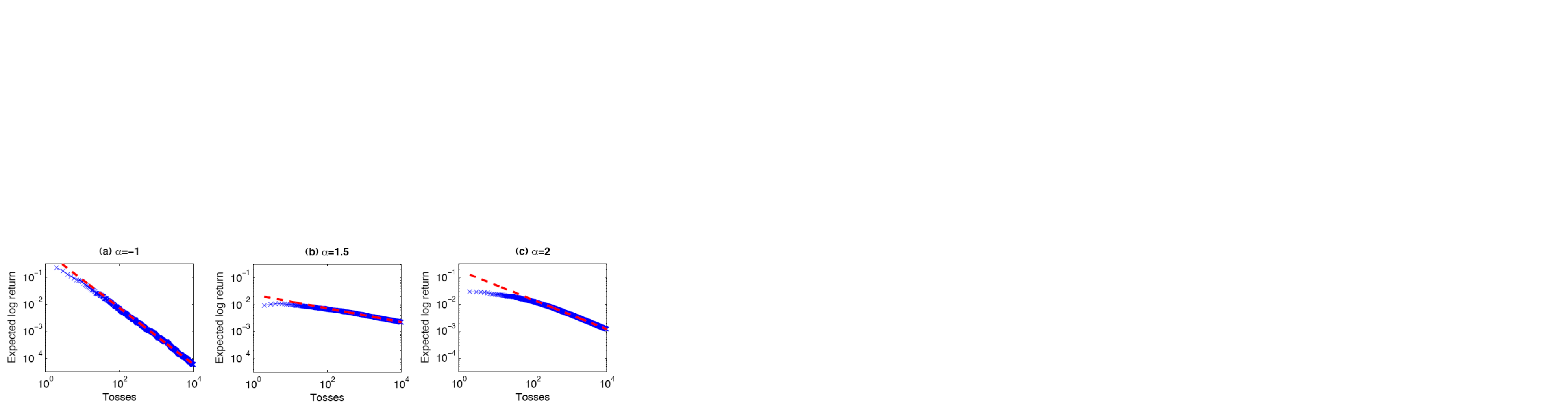}
\caption{The expected log-return to a rational $\epsilon$ player is plotted as a function of time, measured as the number of tosses of the coin.  This provides a measure of the inefficiency of the wealth allocation in the game, which diminishes as the system converges to equilibrium.  A rational $\epsilon$ player is one who follows an optimal strategy but whose total wealth is too small to affect the outcome.  As the game continues the return to an optimal player goes down because the wealth is reallocated to players with better strategies.  In these simulations there are $N=3000$
fixed strategy players whose wealth is initially evenly
distributed. After an initial transient period the expected log
return decreases as a power law in the number of tosses.  Since the exponent is less than one, convergence is extremely slow.  This depends on the degree of
subjectivity of the map: the more subjective (i.e.,
the closer the map is to $q(p) = p$), the slower the
convergence to efficiency. In (a), function $q(p)=1-p$ is used; in (b) and (c)
$q(p)=q_\alpha(p)$ is used, with the values of $\alpha$ specified in the figures.}
\label{efficiency}
\end{center}
\end{figure}

We have studied the approach to efficiency numerically for a variety
of different reality maps, as shown in Figure~\ref{efficiency}.  
To damp out the effect of statistical fluctuations from run to run we
take an ensemble average by varying the random number seed.  For these simulations, in order to get good scaling results across the whole time interval we used $N=3000$, much larger than $N=30$ as used previously\footnote{
The domain of validity of the power law scaling is truncated for small $N$, i.e. for long times the power law scaling breaks down.  The reason for this is apparent from the derivations in Section~\ref{analyticSection}.  If $N$ is small, during the approach to equilibrium all the wealth is concentrated on the winning strategy and its immediate neighbors, and the continuous approximations needed for power law convergence break down.}.
In all but the purely subjective case we find that the
efficiency is a decreasing function of time, asymptotically
converging as a power law $t^{-\gamma}$ with $0 \le \gamma \le 1$.
For the purely objective case $q(p) = 1/2$, or for the self defeating case $q(p) = 1 - p$ we observe $\gamma
\approx 1$.  In other cases we observe $\gamma < 1$, with gamma always in the range
$0 \leq \gamma \leq 1$.  See Table~\ref{exponents} for some examples\footnote{
The observed exponents in this table come from \cite{Miller05}.}.
\begin{table}
\begin{tabular}{|l|c|c|c|c|c|c|} \hline
reality map & $\alpha=2$ & $\alpha = 1.5$ & $\alpha = 0.75$ & $\alpha = 0.5$ & $q(p) = \mbox{const}$ & $q(p) = 1 - p$ \\ \hline
observed $\gamma$ & 0.42 & 0.2 & 0.13 & 0.49 & 0.99 & 1.03 \\ \hline
predicted $\gamma$ & 0.30 & 0.23 & 0.25 & 0.5 &  1 & 1 \\ \hline
\label{table2}
\end{tabular}
\caption{Comparison of numerical vs theoretical predictions of the rate of convergence of the inefficiency to equilibrium, which goes as $t^{-\gamma}$.  The first row specifies the reality map; the first four columns refer to the parameter $\alpha$ in equation~(\ref{qalpha}).  The second row contains the results of numerical simulations, while the third row contains the predictions of the theory.}
\label{exponents}
\end{table}


\subsection{Analytic proof of power law convergence to efficiency}

Our numerical results suggest that the approach to efficiency 
is described by a power law.  We now prove this analytically
by building on the results obtained in Section~\ref{analyticSection}.  

The expected log-return of a player with strategy $s$
was given by equation~(\eqref{maxlogreturn}) above:
$r(s) = q\log\frac{s}{p} + (1-q)\log\frac{(1-s)}{(1-p)}.$
Taking the derivative of $r(s)$ for an $\epsilon$ player, 
we note that, because her wealth is infinitesimal,
neither $p$ nor $q$ depend on her strategy, 
and we have
\begin{equation}
\frac{dr}{ds} = \frac{q}{s} - \frac{1-q}{1-s}.  
\label{drds1}
\end{equation}
Accordingly,
a rational $\epsilon$ player maximizes her average log
return by playing $s=q$.  


In the vicinity of a fixed point $\tilde q = \tilde p$, the fraction
bet on heads can be written $p = \tilde q + (p-\tilde q)
=\tilde q + y$, and the probability of heads can be
written $q = \tilde q + \mu( p- \tilde q) = \tilde q + \mu y$,
where $\mu = q'(\tilde p)$ is the slope of the reality
function at the fixed point, as above.  Written in terms
of $\tilde q, \mu$ and $y$, the expression for the 
expected log-return becomes  
\begin{equation}
r = (\tilde q + \mu y ) \log \frac{\tilde q + \mu y}{\tilde q + y}
+ ( 1 - \tilde q - \mu y ) \log \frac{1- \tilde q - \mu y}{1 -  \tilde q - y}.
\label{logreturny}
\end{equation}
Assuming that we are close to the fixed point (i.e.\ $y$ being small),
we can expand equation~(\eqref{logreturny}) to second order
in $y$ for small $y$.
The expansion takes a different form depending on whether
$\tilde q$ is in the interior of the interval $[0,1]$ or 
at one of the endpoints.  When
$\tilde q$ lies in the interior, $\tilde q \neq 0,1$,
equation~(\eqref{logreturny}) yields
\begin{equation}
r = \frac{1}{2} (1-\mu)^2 \bigg( 
\frac{1}{\tilde q} + \frac{1}{1-\tilde q} \bigg) y^2 + O(y^3)
= \frac{1}{2} (1-\mu)^2 C y^2 + O(y^3),
\label{logreturny2}
\end{equation}
where $C = 1/\tilde q(1-\tilde q)$.

The behavior of the average log-return depends on the
distance from equilibrium in the vicinity of the fixed
point.  This distance is dominated by fluctuations
as described in Section~\ref{analyticSection} above.  
Because these fluctuations obey a power law,
the average log-return also obeys a power law.
The exponent in the power law depends on the slope
of the reality map at the fixed point, and on whether
the fixed point is in the interior of the interval or whether it is one of the boundary points at $0$ or $1$. 
For comparison to the numerical results it is useful to note that for the reality map of equation~(\ref{qalpha}) the slope $\mu = q'(\tilde{p}) = \alpha$ for the fixed point at $p = 1/2$ and $\mu = 8/(\pi^2\alpha)$ for the fixed points at $p = 0$ or $p = 1$.  We also remind the reader that we do not expect our estimates of $\gamma$ to be accurate when the fixed point is in the interior and $\mu \approx 1$ or when it is on the boundary and $\mu \approx 0$.   

Depending on the reality map, the log-return of the rational $\epsilon$ player approaches zero as $t ^{-\gamma}$ as follows.

\begin{itemize}
\item
\textit{If $\tilde q \neq 0,1$,
and $\mu < 0$ then $\gamma = 1$.}  The average distance from  
equilibrium in this case goes as $\Delta_t
 \propto t^{-1/2}$.
Accordingly the average value of $y^2$ in equation~(\eqref{logreturny2})
goes as $\propto t^{-1}$:
the average log-return obeys power law with exponent
$-1$.  This is true, for example, for the objective case $q(p) = \mbox{const}$ and the self-defeating case $q(p) = 1 - p$.  (See Table~\ref{exponents} and Figure~\ref{efficiency}(a)).
\item
\textit{If $\tilde q \neq 0,1$ and $\mu  \geq 0$ then $\gamma =  (1 - \mu)$.}
The average value of $y^2$ in this case is dominated
by $ \Delta_t^2 \propto t^{(\mu-1)}$.  This is the case for $\alpha = 0.5$ and $\alpha = 0.75$ in Table~\ref{exponents}.  For $\alpha = 0.5$ the agreement is quite good, but as expected, for $\alpha = 0.75$, where $\mu = 0.75$, we are closer to the critical value $\mu = 1$ and the agreement is not as good.
\item
\textit{If $\tilde q=0$ or $\tilde q = 1$ then $\gamma = (1 - \mu)/2$.}
In this case the expansion in small $y$ yields
\begin{equation}
r = (1-\mu + \mu \log \mu) y + O(y^2).
\label{logreturny3}
\end{equation}
Once again, the typical distance from the fixed point goes
as the average fluctuation size, $\Delta_t \propto t^{(\mu-1)/2}$,
and so does the expected log-return
for a rational $\epsilon$ player.  SeeTable~\ref{exponents} with $\alpha = 1.5$ and $\alpha = 2$ and Figure~\ref{efficiency}(b) and (c).  Agreement is good for $\alpha = 1.5$, but as expected for $\alpha = 2$, where $\mu$ is closer to zero, the agreement is not as good.
\end{itemize}

To summarize, for self-defeating or purely objective reality maps $\gamma = 1$.  For self-reinforcing reality maps, as we approach the purely subjective case where $\mu = 1$, $\gamma \to 0$.  Thus, the converge of the inefficiency to zero is always slow, but when the reality map is strongly subjective, it is extremely slow.

\section{Summary}

We have introduced a very simple evolutionary game of chance with the nice property that one can explicitly study the influence of
the player's actions on the outcome of the game.  By altering the
reality map $q(p)$ it is possible to continuously vary the setting
from completely objective, i.e.\ the odds are independent of the
players' actions, to completely subjective, i.e.\ the odds are
completely determined by the players' actions.

It has long been known that subjective effects can play an important
role in games, causing problems such as increasing returns and
lock-in to a particular outcome due to chance events.  This game illustrates these effects nicely, making it possible to study both positive and negative feedback.

Perhaps the nicest feature of this game is that it provides a setting in which to study the progression from an inefficient to an efficient market.  We have
introduced a method of measuring the inefficiency of the game when it is out of equilibrium that is similar to how one might quantitatively measure the magnitude of the arbitrage opportunities for an optimal player in a financial market.  As time goes on the game tends to become more efficient, with the inefficiency dying out as a power law of the form $t^{-\gamma}$.  As shown in the previous section, $\gamma$ is always in the range $0 \leq \gamma \leq 1$.  For self-defeating or purely objective reality maps $\gamma = 1$.  This already implies a very slow convergence to efficiency.  When the reality map is close to being subjective, however, $\gamma \to 0$, implying {\it extremely} slow convergence.

One might consider several extensions of the problem studied here.
For example, one could study learning (see e.g.\ \cite{sato-akiyama-farmer:PNAS}).
Another interesting possibility is to allow more general reality
maps, in which $q$ is a multidimensional function with a
multidimensional argument that may depend on the bets of individual
players.  For example, an interesting case is to allow some players,
who might be called pundits, to have more influence on the outcome
than others.  It would also be very interesting to modify the game
so that it is an open system, e.g.\ relaxing the wealth conservation
condition and allowing external inputs.  This may prevent the
asymptotic convergence of all the wealth to a single player,
creating more interesting long-term dynamics.


\bibliographystyle{jedc}
\bibliography{jdf,pnas}
\end{document}